\documentclass[prb,preprint,eqsecnum,aps]{revtex4}
\usepackage{graphicx}
\usepackage{epsf}

\newcommand{\e}{{\rm e}}
\newcommand{\ep}{\epsilon}

\newcommand{\dte}{{\epsilon}}

\newcommand{\x}{{\bf x}}

\newcommand{\br}{{\bf r}}

\newcommand{\sgn}{{\rm sgn}}

\newcommand{\bea}{\begin{eqnarray}}
\newcommand{\eea}{\end{eqnarray}}
\newcommand{\be}{\begin{equation}}
\newcommand{\ee}{\end{equation}}
\newcommand{\ba}{\begin{eqnarray}}
\newcommand{\ea}{\end{eqnarray}}

\newcommand{\nn}{\nonumber}
\newcommand{\la}{\label}

\def\t1{e_{_T}}
\def\v1{e_{_V}}

\begin{document}
\title{Simple proof that there is no sign problem in Path Integral Monte Carlo simulations of fermions
in one dimension}

\author{Siu A. Chin}
\email{chin@physics.tamu.edu.}

\affiliation{Department of Physics and Astronomy,
Texas A\&M University, College Station, TX 77843, USA}

\begin{abstract}

It is widely known that there is no sign problem in Path Integral Monte Carlo (PIMC) simulations of fermions
in one dimension. Yet, as far as the author is aware, there is no direct proof of this in the literature. 
This work shows that the {\it sign} of the $N$-fermion anti-symmetric free propagator
is given by the product of all possible pairs of particle separations, or relative displacements. 
For a non-vanishing closed-loop product of such propagators, as required by PIMC,
all relative displacements from adjacent propagators 
are paired into perfect squares, and therefore the loop product must be positive, 
but only in one dimension. By comparison, 
permutation sampling, which does not evaluate the determinant of
the anti-symmetric propagator exactly, remains
plagued by a low-level sign problem, even in one dimension.

\end{abstract}
\maketitle

\section {Introductions}

It is common knowledge since Takahashi and Imada's
calculation\cite{tak84} that there is no sign problem in
Path-Integral Monte Carlo (PIMC) simulations of fermions in one dimension.
This conclusion is strongly supported by theoretical arguments by Girardeau\cite{gir60}, 
Negele and Orland\cite{neg88}, and Ceperley\cite{cep91}, but none of them are
actual proofs specifically for PIMC. Only recently has the author given a proof\cite{chin23} of this, 
based on Girardeau's\cite{gir65} topological insight. 

That topological idea is very simple in the case of two fermions
in one dimension with coordinates $x_1$ and $x_2$. Since the propagator must change sign when
the two positions are exchanged, the positive and negative regions of the propagator are 
completely separated by the nodal line $x_1=x_2$ where the propagator vanishes.
A closed-loop product of propagators in 
the plane of $(x_1,x_2)$, as required in PIMC,
must therefore cross this infinite nodal line, and changes sign, either zero, or even number of times. 
Hence, the sign of a non-vanishing closed loop product of propagators must be positive, 
with no sign problem. 

At higher dimensions,
say two-dimension, the space of the propagator is four dimensional $(x_1,y_1,x_2,y_2)$, 
but the {\it coincidental} nodal plane\cite{cep91}
given by $x_1=x_2$ and $y_1=y_2$ is only two dimensional. 
Just as a line, which is two dimensions less, cannot divide the three-dimensional space,
a two-dimensional nodal plane, also cannot divide the four-dimensional space
into two halves.
Hence, the previous argument fails and there is a sign problem in more than one dimension.

Since this topological argument for the existence of the sign problem in more than one
dimension is less intuitive, I alternately determined the sign
of the two-fermion propagator directly in terms of their relative displacements \cite{chin23}.
Since only in one dimension can relative displacements from adjacent propagators
be paired into pure squares, the sign problem is absent only in one dimension.
For completeness, this important two fermion case for understanding 
the dimensional dependence of the sign problem is restated in Sect.III below. 
However, at the time of Ref.\onlinecite{chin23}'s publication,
there was no known way of determining the sign of an arbitrary $N$-fermion propagator in one dimension, 
and hence no general proof by direct sign determination.  

This work, by use of Mikhailov's expansion\cite{mik01}
in terms of Vandermonde determinants\cite{vei06}, can now compute
the sign of the $N$-fermion propagator directly and present a much simpler proof.
This work, which {\it only computes the sign} of the fermion propagator, is purely a technical achievement, filling a missing gap in the literature.
However, as discussed in the Conclusion, this proof can now explain why
permutation sampling\cite{cep95,lyu05}, which does not
evaluate the fermion propagator's determinant completely, 
remains plagued by a low-level sign problem\cite{lyu05}, even in one dimension. 

This work will be concise in presenting only technical details, 
Ref.\onlinecite{chin23} can be consulted for more background discussions. 
After a brief summary of key PIMC equations and defining the sign problem in Sect.\ref{fpimc},
Sect.\ref{nosign} answers the frequently asked question of why no sign problem only in one dimension.
The sign of two, three and $N$-fermion propagators
is then determined in successive sections \ref{twofer}-\ref{nfer}.
A concluding summary is given in Sect.\ref{con}, with a comparative discussion on permutation
sampling.

\section {Fermion Path Integral Monte Carlo}
\la{fpimc}

Let $\x=(\br_1,\br_2 \cdots \br_N)$ denote the coordinates of $N$ fermions in $d$-dimension.
At the heart of PIMC is the Monte Carlo sampling of the closed-end, $k$-bead
path integral
\ba
G_{k}(\x,\x;\tau)&=&\langle \x|(\e^{-\dte(\hat T+\hat V)})^k|\x\rangle\nn\\
&=&\int_{-\infty}^{\infty} d\x_1\cdots d\x_{k-1}\,
G_1(\x,\x_1;\dte)G_1(\x_1,\x_2;\dte)\cdots G_1(\x_{k-1},\x;\dte)
\la{mb}
\ea
at imaginary time $\tau=k\dte$, where $\hat T$ and $\hat V$ are the kinetic and potential
operators of the many-fermion system, and $G_1(\x',\x;\dte)$ is a short-time propagator,
of which the simplest is the primitive second-order approximation
\ba
G_1(\x',\x;\dte)
&=&\langle \x'|\e^{-\dte(\hat T+\hat V)}|\x\rangle\nn\\
&\approx&\e^{-(\dte/2) V(\x')}
G_0(\x',\x;\dte)
\e^{-(\dte/2) V(\x)},
\la{pap}
\ea
where $G_0(\x',\x;\dte)$ is the anti-symmetric free-fermion propagator
\ba
G_0(\x',\x;\dte)&=&\frac1{N!}{\rm det}\left(\frac1{(2\pi\dte)^{d/2}}
\exp\left[-\frac1{2\dte}(\br_i^\prime-\br_j)^2\right] \right).
\la{free}
\ea

Since $G_0(\x',\x;\dte)$ is not positive definite, the integrand in the loop-integral (\ref{mb}) can be negative
for some paths. This is the fermion sign problem in PIMC. The goal of this work is to show that,
despite the fact that
$G_0(\x',\x;\dte)$ can be of either sign, {\it the integrand of (\ref{mb})
over any  closed path}
\be \x\rightarrow \x_1\rightarrow \x_2\rightarrow\cdots \x_{k-1}\rightarrow \x,
\ee
{\it if it is non-vanishing, is always positive in one dimension}. 

The sign problem is due entirely to the fact that the free-fermion propagator (\ref{free}) can be negative.
The interacting potential $V(\x)$ is exponentiated in (\ref{pap}) and the exponential function
 is always positive regardless whether the potential is attractive or repulsive. In one dimension, where
there is no sign problem, the potential has no effect on  
sign of the integrand in (\ref{mb}). 
In more than one dimension, where there is a sign problem, it is possible  that
the exponentiated potential in (\ref{pap})  can further aggravating the sign problem
by giving more weight to the negative region of the integrand.

\section {Why no sign problem only in one dimension}
\la{nosign}

In $d$-dimension, the two-fermion free propagator from (\ref{free}) is given by
\ba
G_0(\br_1^\prime,\br_2^\prime,\br_1,\br_2;\dte)
&=&\frac12\frac1{(2\pi\dte)^d}
\det\left(\begin{array}{cc}
	\e^{-(\br'_1-\br_1)^2/(2\dte)} &  \e^{-(\br'_1-\br_2)^2/(2\dte)}\\
	\e^{-(\br'_2-\br_1)^2/(2\dte)} & \e^{-(\br'_2-\br_2)^2/(2\dte)}
     \end{array}\right).\la{twofd}\\
&=&\frac12\frac1{(2\pi\dte)^d}\e^{-\frac1{2\dte}\left[(\br_1^\prime-\br_1)^2
	+(\br_2^\prime-\br_2)^2\right] }
\left(1
-\e^{-\frac1{\dte}(\br_2^\prime-\br^\prime_1)\cdot(\br_2-\br_1) } \right),
\la{propr2}
\ea
whose sign of is determined by the sign of
\be
1-\exp(-\frac1{\dte}\br_{21}'\cdot\br_{21}),
\ee
where $\br_{21}'=\br_2'-\br_1'$ and $\br_{21}=\br_2-\br_1$, which in turn is
given by the sign of $\br_{21}'\cdot\br_{21}$. Therefore one has
\be
\sgn\Bigl(G_0(\br_1',\br_2',\br_1,\br_2;\dte)\Bigr)
=\sgn(\br_{21}'\cdot\br_{21}).
\ee
Note that the sign of the propagator is solely determined by particle positions and is
independent of $\dte$.

Since the trace of one and two propagators are always positive, the sign problem only appears
for three or more propagators:
\ba
\sgn\Bigl(G_0(\br_{21},\br_{21}')G_0(\br_{21}',\br_{21}'') G_0(\br_{21}'',\br_{21})\Bigr)
&=&\sgn\Bigl((\br_{21}\cdot\br_{21}')(\br_{21}'\cdot\br_{21}'')(\br_{21}''\cdot\br_{21})\Bigr),\nn\\
&=&|\br_{21}|^2|\br_{21}'|^2|\br_{21}''|^2\sgn\Bigl(\cos\theta\cos\theta'\cos\theta''\Bigr).
\ea
Since the cosine functions can take both signs, the sign problems exist whenever the dot product
produces a cosine function, {\it i.e.}, at dimensions greater than one. At one dimension, there's no angles,
no cosine functions and the sign is just
\ba
\sgn\Bigl(G_0(x_{21},x_{21}')G_0(x_{21}',x_{21}'') G_0(x_{21}'',x_{21})\Bigr)
&=&\sgn\Bigl((x_{21}x_{21}')(x_{21}'x_{21}'')(x_{21}'' x_{21})\Bigr),\nn\\
&=&\sgn\Bigl((x_{21})^2(x_{21}')^2(x_{21}'')^2\Bigr)\ge 0,
\ea
where all displacements have paired up as perfect squares, and hence no sign problem.
The above can clearly be generalized to a loop any number of propagators.
To prove the general case, one only needs to determine the sign of the $N$-fermion propagator
in one dimenison.

\section {The sign of the two-fermion propagator}
\la{twofer}

The method of computing the sign of the propagator by directly evaluating the determinant
cannot be easily generalized to more than two fermions. Here, we first cast the one dimensional 
propagator into a form suggested in Ref.\onlinecite{chin23}.  

For two (spinless) fermions, the one dimensional form of (\ref{twofd}) is just
\ba
G_0(x_1^\prime,x_2^\prime,x_1,x_2;\dte)
&=&\frac12\frac1{2\pi\dte}
\det\left(\begin{array}{cc}
	\e^{-(x'_1-x_1)^2/(2\dte)} &  \e^{-(x'_1-x_2)^2/(2\dte)}\\
	\e^{-(x'_2-x_1)^2/(2\dte)} & \e^{-(x'_2-x_2)^2/(2\dte)}
     \end{array}\right).
\la{ftwo}     
\ea
Since this work is not interested in the actual value of the propagator, but
only its {\it sign},  all normalization
factors and purely positive functions can be ignored. 

By factoring out, from (\ref{ftwo}),  $\e^{-x'^2_1/(2\dte)}$, $\e^{-x'^2_2/(2\dte)}$
from row one and two respectively, and $\e^{-x^2_1/(2\dte)}$,  $\e^{-x^2_2/(2\dte)}$
from column one and two, one has
\ba
G_0(x_1^\prime,x_2^\prime,x_1,x_2;\dte)
&\propto&	\e^{-(x'^2_1+x'^2_2+x^2_1+x^2_2)/(2\dte)}
\det\left(\begin{array}{cc}
	\e^{x'_1x_1/\dte} & \e^{x'_1x_2/\dte}\\
	\e^{x'_2x_1/\dte} & \e^{x'_2x_2/\dte}
\end{array}\right).
\la{twof}
\ea
The sign of $G_0$ is therefore just the sign of the above determinant. 
As noted in Sect.{\ref{nosign},
since the sign is fixed by the positions only, independent 
of $\dte$, it can be determined in the limit of $\dte\rightarrow\infty$, yielding
\ba
\sgn(G_0(x_1^\prime,x_2^\prime,x_1,x_2;\dte))
&=&\sgn( (x'_1x_1+x'_2x_2-x'_1x_2-x'_2x_1)/\dte)\nn\\
&=&\sgn( x'_{21}x_{21} ).
\la{fdet}
\ea
It then follows that for $\x=(x_1,x_2)$, the sign of a loop product of $n$ propagator  is given by
\ba
&&\sgn(G_0(\x,\x^{\prime};\dte)G_0(\x^{\prime},\x^{\prime\prime};\dte)
G_0(\x^{\prime\prime},\x^{\prime\prime\prime};\dte)\cdots G_0(\x^{\{n-1\}},\x;\dte))\nn\\
&&\qquad\qquad=\sgn(x_{21}x_{21}^{\prime}x_{21}^{\prime}x_{21}^{\prime\prime}x_{21}^{\prime\prime}
\cdots  x_{21}^{\{n-1\}}x_{21})\la{abo}\\
&&\qquad\qquad=\sgn( (x_{21})^2(x^\prime_{21})^2(x^{\prime\prime}_{21})^2\cdots(x^{\{n-1\}}_{21})^2)\ge 0.
\la{nonep}
\ea
Thus for a product of any number of two-fermion propagators, if it is non-vanishing, then its sign must
be positive because for a closed loop, relative displacements from adjacent propagators
will always pair up to  a perfect square.

The determinant in (\ref{twof}) can now be evaluated by an alternative 
method generalizable to $N$ fermions.
Since the $\ep\rightarrow\infty$ limit is the same as the $x,x'\rightarrow 0$ limit, one can just do the latter and
suppress the appearance of $\ep$. 
For notational clarity, we will also replace $x'$ by $s$ in the discussion below
and evaluate the determinant as follow:
\ba
\det\left(\begin{array}{cc}
	\e^{x_1 s_1} & \e^{x_1 s_2}\\
	\e^{x_2 s_1} & \e^{x_2 s_2}
\end{array}\right)&=&\e^{x_1 s_1}\e^{x_2 s_2}-\e^{x_2 s_1}\e^{x_1 s_2}\nn\\
&=&\sum_{n_1=0}^\infty\sum_{n_2=0}^\infty\frac1{n_1!n_2!}
(x_1^{n_1} s_1^{n_1}x_2^{n_2} s_2^{n_2}-x_2^{n_1}s_1^{n_1}x_1^{n_2} s_2^{n_2})\nn\\
&=&\sum_{n_1=0}^\infty\sum_{n_2=0}^\infty\frac1{n_1!n_2!}s_1^{n_1}s_2^{n_2}
(x_1^{n_1}x_2^{n_2}-x_2^{n_1}x_1^{n_2})\nn\\
&=&\sum_{n_1=0}^\infty\sum_{n_2=0}^\infty\frac1{n_1!n_2!}s_1^{n_1}s_2^{n_2}
\det\left(\begin{array}{cc}
	x_1^{n_1} & x_1^{n_2}\\
	x_2^{n_1} & x_2^{n_2} 
\end{array}\right).
\la{tfdet}
\ea
Since $n_i$ serve as a column index, the determinant above vanishes for
$n_1=n_2$. Therefore the sum is over $n_1<n_2$ and $n_2<n_1$ only. 
The latter case can be viewed as the former case with $n_1$ interchanged with $n_2$.
This changes the column of the determinant, corresponding to the original determinant with a negative sign, hence
\ba
\det\left(\begin{array}{cc}
	\e^{x_1 s_1} & \e^{x_1 s_2}\\
	\e^{x_2 s_1} & \e^{x_2 s_2}
\end{array}\right)
&=&\sum_{n_1<n_2}\frac1{n_1!n_2!}\Biggl[s_1^{n_1}s_2^{n_2}
\det\left(\begin{array}{cc}
	x_1^{n_1} & x_1^{n_2}\\
	x_2^{n_1} & x_2^{n_2}
\end{array}\right)
-s_1^{n_2}s_2^{n_1}
\det\left(\begin{array}{cc}
	x_1^{n_1} & x_1^{n_2}\\
	x_2^{n_1} & x_2^{n_2}
\end{array}\right)\Biggr]
\nn\\
&=&\sum_{n_1<n_2}\frac1{n_1!n_2!}(s_1^{n_1}s_2^{n_2}-s_1^{n_2}s_2^{n_1})
\det\left(\begin{array}{cc}
	x_1^{n_1} & x_1^{n_2}\\
	x_2^{n_1} & x_2^{n_2}
\end{array}\right)
\nn\\
&=&\sum_{n_1<n_2}\frac1{n_1!n_2!}
\det\left(\begin{array}{cc}
	s_1^{n_1} & s_1^{n_2}\\
	s_2^{n_1} & s_2^{n_2}
\end{array}\right)
\det\left(\begin{array}{cc}
	x_1^{n_1} & x_1^{n_2}\\
	x_2^{n_1} & x_2^{n_2}
\end{array}\right).
\la{detp}
\ea
The above is the simplest $2\times 2$ version of of Mikhailov's method\cite{mik01} of
expanding a determinant of mixed variable into a product of two determinants of
separated variables.

In the limit of $s_i,x_i\rightarrow 0$, the single leading order term in the above sum is given by $n_1=0$ and $n_2=1$: 
\ba
\det\left(\begin{array}{cc}
	\e^{x_1 s_1} & \e^{x_1 s_2}\\
	\e^{x_2 s_1} & \e^{x_2 s_2}
\end{array}\right)
\rightarrow
\det\left(\begin{array}{cc}
	1 & s_1\\
	1 & s_2
\end{array}\right)
\det\left(\begin{array}{cc}
	1 & x_1\\
	1 & x_2
\end{array}\right)=(s_2-s_1)(x_2-x_1)=s_{21}x_{21},
\la{tmat}
\ea
reproducing the sign of the two-fermion propagator (\ref{fdet}).

\section {The sign of the three-fermion propagator}
\la{threefer}

For three fermions, 
\ba
&&\det\left(\begin{array}{ccc}
	\e^{x_1 s_1} & \e^{x_1 s_2}& \e^{x_1 s_3}\\
	\e^{x_2 s_1} & \e^{x_2 s_2}& \e^{x_2 s_3}\\
	\e^{x_3 s_1} & \e^{x_3 s_2}& \e^{x_3 s_3}
\end{array}\right)\nn\\
&&=
\det\left(\begin{array}{cc}
	\e^{x_1 s_1}& \e^{x_1 s_2}\\
	\e^{x_2 s_1}& \e^{x_2 s_2}
\end{array}\right)\e^{x_3 s_3}
-
\det\left(\begin{array}{ccc}
	\e^{x_1 s_1} & \e^{x_1 s_2}\\
	\e^{x_3 s_1} & \e^{x_3 s_2}
\end{array}\right)\e^{x_2 s_3}
+
\det\left(\begin{array}{ccc}
	\e^{x_2 s_1} & \e^{x_2 s_2}\\
	\e^{x_3 s_1} & \e^{x_3 s_2}
\end{array}\right)\e^{x_1 s_3}\nn\\
&&=
\det\left(\begin{array}{cc}
	\e^{x_1 s_1}& \e^{x_1 s_2}\\
	\e^{x_2 s_1}& \e^{x_2 s_2}
\end{array}\right)\e^{x_3 s_3}
-(x_2\leftrightarrow x_3)+(x_2\leftrightarrow x_3\ {\rm then}\ x_1\leftrightarrow x_2),
\ea
corresponding to
\ba
&&=\sum_{n_1=0}^\infty\sum_{n_2=0}^\infty\sum_{n_3=0}^\infty\frac1{n_1!n_2!n_3!}s_1^{n_1}s_2^{n_2}s_3^{n_3}
\Bigl[\ (x_1^{n_1}x_2^{n_2}-x_2^{n_1}x_1^{n_2}) x_3^{n_3}\nn\\
&&\qquad\qquad\qquad\qquad\qquad\qquad\quad\qquad -(x_1^{n_1}x_3^{n_2}-x_3^{n_1}x_1^{n_2})x_2^{n_3}\nn\\
&&\qquad\qquad\qquad\qquad\qquad\qquad\quad\qquad +(x_2^{n_1}x_3^{n_2}-x_3^{n_1}x_2^{n_2})x_1^{n_3}\Bigr]\nn\\
&&=\sum_{n_1=0}^\infty\sum_{n_2=0}^\infty\sum_{n_3=0}^\infty
\frac1{n_1!n_2!n_3!}s_1^{n_1}s_2^{n_2}s_3^{n_3}
\det\left(\begin{array}{ccc}
	x_1^{n_1} & x_1^{n_2}& x_1^{n_3}\\
	x_2^{n_1} & x_2^{n_2}& x_2^{n_3}\\
	x_3^{n_1} & x_3^{n_2}& x_3^{n_3}
\end{array}\right).
\ea
This is the three fermion generalization of (\ref{tfdet}).
As in the two-fermion case, the sign changes in permuting the unrestricted sum $\sum_{n_1,n_2,n_3}$
into the ordered sum $\sum_{n_1<n_2<n_3}$ result in a determinant for $s_i$:
\ba
\det\left(\begin{array}{ccc}
	\e^{x_1 s_1} & \e^{x_1 s_2}& \e^{x_1 s_3}\\
	\e^{x_2 s_1} & \e^{x_2 s_2}& \e^{x_2 s_3}\\
	\e^{x_3 s_1} & \e^{x_3 s_2}& \e^{x_3 s_3}
\end{array}\right)
=
\sum_{n_1<n_2<n_3}^\infty\frac1{n_1!n_2!n_3!}
\det\left(\begin{array}{ccc}
	s_1^{n_1} & s_1^{n_2}& s_1^{n_3}\\
	s_2^{n_1} & s_2^{n_2}& s_2^{n_3}\\
	s_3^{n_1} & s_3^{n_2}& s_3^{n_3}
\end{array}\right)
\det\left(\begin{array}{ccc}
	x_1^{n_1} & x_1^{n_2}& x_1^{n_3}\\
	x_2^{n_1} & x_2^{n_2}& x_2^{n_3}\\
	x_3^{n_1} & x_3^{n_2}& x_3^{n_3}
\end{array}\right).\ \ \ \
\la{three}
\ea
This is the $3\times 3$ version of Mikhailov's method\cite{mik01} of 
expanding a mixed variable determinant.

The leading order term in the above sum is $n_1=0$, $n_2=1$ and $n_3=2$, 
\ba
&&\rightarrow\frac1{2}
\det\left(\begin{array}{ccc}
	1 & s_1& s_1^{2}\\
	1 & s_2& s_2^{2}\\
	1 & s_3& s_3^{2}
\end{array}\right)
\det\left(\begin{array}{ccc}
	1 & x_1& x_1^{2}\\
	1 & x_2& x_2^{2}\\
	1 & x_3& x_3^{2}
\end{array}\right)
=\frac12 s_{21}s_{31}s_{32} x_{21}x_{31}x_{32},
\la{thmat}
\ea
which correctly changes sign whenever any pair of particles is exchanged.
All relative displacements will again pair up as perfect squares for any closed loop
product of propagators.

\section {The sign of the N-fermion propagator}
\la{nfer}

The two determinants in (\ref{tmat}) and (\ref{thmat})
are simple cases of the general $N\times N$ Vandermonde determinant\cite{vei06}:
\be
\det\left(\begin{array}{ccccc}
	1 & x_1 &x_1^2& \cdots &x_1^{N-1}\\
	1 & x_2 &x_2^2& \cdots &x_2^{N-1}\\
	1 & x_2 &x_2^2& \cdots &x_2^{N-1}\\		
	1 & \cdots &\cdots& \cdots &\cdots\\	
	1 & x_N &x_n^2& \cdots &x_N^{N-1}		
\end{array}\right)=\prod_{1\le i<j\le N}(x_j-x_i)
\la{vand}
\ee
The generalization of (\ref{thmat}) to 
$N$ fermions then follows from (\ref{vand}) immediately as
\be
G_0(s_1,s_2,\cdots s_N,x_1,x_2,\cdots x_N,\dte)\propto
\Biggl(\prod_{i<j}s_{ji}\Biggr)\Biggl(\prod_{i<j} x_{ji}\Biggr),
\la{nprog}
\ee
which changes sign whenever any pair of $s_i$ or $x_i$ is exchanged.
This then again entails that a non-vanishing closed-loop product of propagators (\ref{nprog}) will
have paired displacements as squares and its sign will always be positive.

\section {Conclusions}
\la{con}

This work has proved that fermion PIMC using anti-symmetric free propagators has no sign problem
in one dimension. This proof is based solely on determining the sign of the $N$-fermion free propagator,
without referencing anything external, such as equivalent
bosons\cite{gir60}, a preferred subspace\cite{neg88}, restricted nodal regions\cite{cep91} or even 
topology\cite{chin23}. The proof gives insight 
into why there is no sign problem in one dimension by showing that in a closed loop of propagators,
all relative displacements are paired up as squares.
However, the proof depends crucially on knowing the sign of the determinant exactly, as dictated by (\ref{nprog}).
This implies that if fermion PIMC is implemented by sampling permutations\cite{cep95,lyu05} only,  
then the determinant's sign is not exactly determined, with
displacements not precisely paired up as squares. 
Permutation sampling (PS) refers to the fact that 
the anti-symmetric free-fermion propagator (\ref{free}) 
can be expanded as a sum over permutations, 
\ba
G_0(\x',\x;\dte)&=&\frac1{N!}{\rm det}\left(\frac1{\sqrt{2\pi\dte}}
\exp\left[-\frac1{2\dte}(x_i^\prime-x_j)^2\right] \right),\nn\\
&=&\frac1{N!}\sum_{P}(-1)^P \frac1{(2\pi\dte)^{N/2}}\exp\left[-\frac1{2\dte}(\x^\prime-\x_P)^2\right],
\ea
but only a single random permutation is sampled for each propagator.
This clearly then does not
determine the sign of each propagator exactly but only on the average.   
The result is a lingering low-level sign problem,
yielding poorer results than that of evaluating the determinant  of the anti-symmetric propagator (AP), as
reported by Lyubartsev\cite{lyu05}:
\begin{quotation}
``In the case of the PS scheme, it was only possible to evaluate the density of the
 first excited state of the one-dimensional harmonic oscillator with a few per cent precision.
 However, in the case of the AP scheme applied to the same system, very accurate estimations
 of the densities of up to at least the eighth excited state become possible. The key to success
 lies in the fact that the AP scheme solves the sign problem completely, providing a strictly
 positive weight function for fermions in one dimension."
\end{quotation}
The fundamental reason why the AP scheme should provide 
``a strictly positive weight function for fermions in one dimension" 
can now be understood, according to this work, as due to
the fact that all relative displacements from adjacent propagators can be paired up as pure 
squares only in one dimension.

\end{document}